\documentclass[twocolumn,prl,aps,notitlepage]{revtex4}
\usepackage{tikz}
\usepackage{amssymb}
\usepackage{psfrag}
\usepackage{textcomp}
\usepackage{pdfsync}
\usepackage{amsmath}
\usepackage{amsfonts}
\usepackage{graphicx,bm,epstopdf}
\usepackage{subfigure}
\usepackage{comment}
\usepackage{verbatim}
\usepackage{color}
\usepackage{upgreek}
\usepackage{natbib,hyperref}
\usepackage{threeparttable}
\usepackage{ifpdf}

\begin{document}

\title{Nonlocal Anomalous Hall Effect}
\author{{Steven S.-L. Zhang and Giovanni Vignale}}
\affiliation{{\ Department of Physics and Astronomy, University of Missouri, Columbia, MO
65211}}
\date{\today }

\begin{abstract}
The anomalous Hall effect is deemed to be a unique transport property of
ferromagnetic metals, caused by the concerted action of spin polarization
and spin-orbit coupling. Nevertheless, recent experiments have shown that
the effect also occurs in a nonmagnetic metal (Pt) in contact with a
magnetic insulator (yttrium iron garnet (YIG)), even when precautions are
taken to ensure there is no induced magnetization in the metal. We propose a
theory of this effect based on the combined action of spin-dependent
scattering from the magnetic interface and the spin Hall effect in the bulk
of the metal. At variance with previous theories, we predict the effect to
be of first order in the spin-orbit coupling, just as the conventional
anomalous Hall effect -- the only difference being the spatial separation of
the spin orbit interaction and the magnetization. For this reason we name
this effect \textit{nonlocal anomalous Hall effect} and predict that its
sign will be determined by the sign of the spin Hall angle in the metal. The
AH conductivity that we calculate from our theory is in good agreement with
the measured values in Pt/YIG structures.
\end{abstract}

\maketitle

\textit{Introduction.}$-$ The anomalous Hall (AH) effect is the generation
of an electric current perpendicular to the electric field in a
ferromagnetic metal~\cite{Nagaosa10RMP}. At variance with the ordinary Hall
effect, which arises from the action of a magnetic field on the orbital

motion of the electrons, the AH effect is ascribed to strong spin-orbit
coupling in concert with spin-polarized itinerant electrons. The spin orbit
coupling plays a central role in inducing a left-right asymmetry (with
respect to the direction of the electric field) in the scattering of
electrons of opposite spins. It is this asymmetry that generates a
transverse charge current from a longitudinal spin current. The same
scattering process generates a pure transverse spin current for systems with
spin unpolarized electrons, which is known as spin Hall effect~\cite%
{Hirsch99PRL,sZhang00PRL,Vignaleg10JSNM, sinova14arxiv_SHreview}. Based on
this picture, the conventional AH effect appears at \textit{first order} in
spin orbit coupling, no matter which kind of microscopic mechanisms
predominates.
\begin{figure*}[tph]
\centering \hspace*{\fill}
\subfigure[~Spin Hall AH effect ($\propto\theta_{sh}^2$)] {
\includegraphics[trim={5.861cm 0.643cm 4.086cm 0.412cm},clip=true, width=0.38\linewidth]{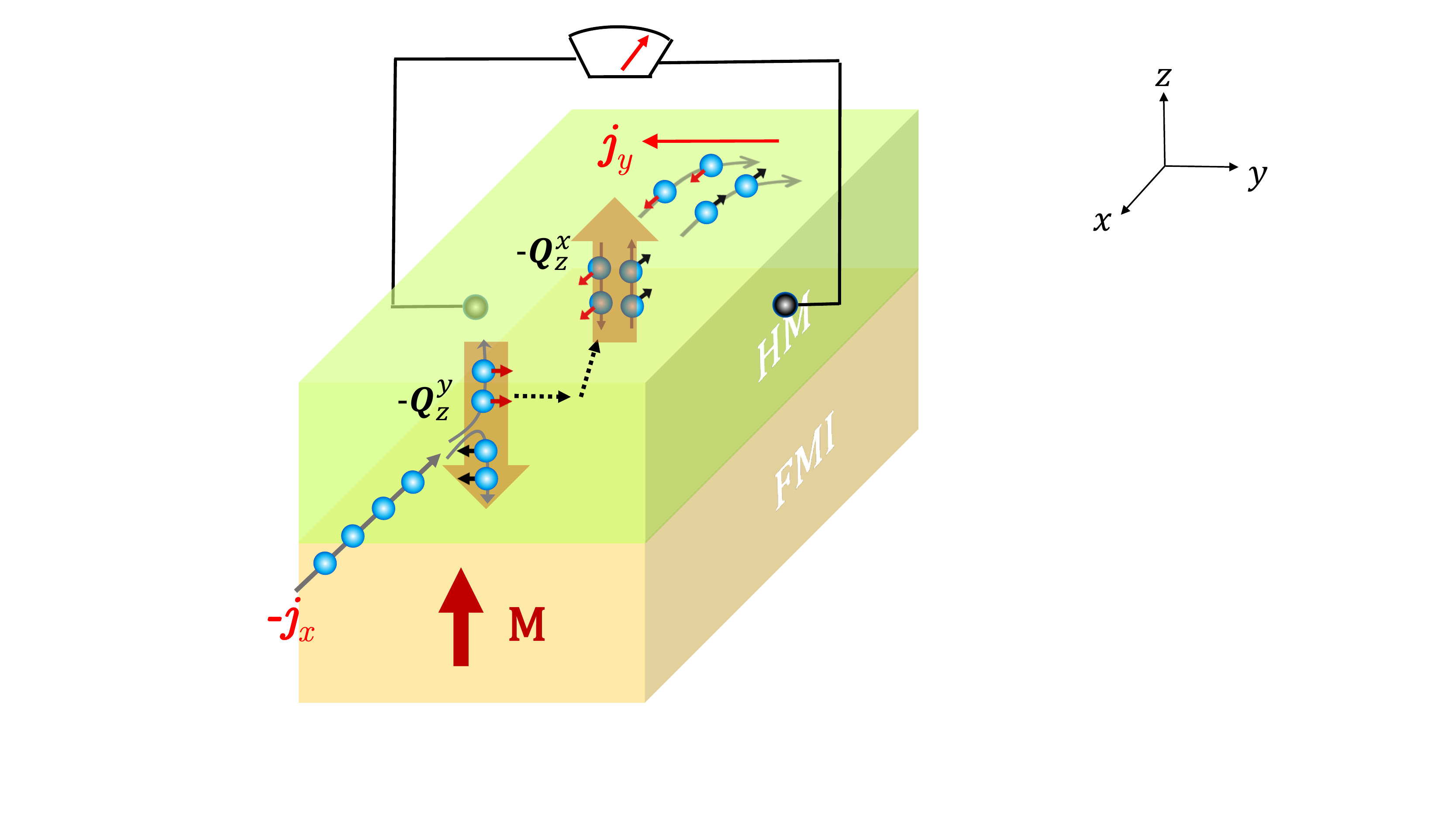}}
\quad
\subfigure[~Nonlocal AH effect ($\propto \theta_{sh}$)] {
\includegraphics[trim={0.95cm 1.055cm 0.19cm 0.317cm},clip=true, width=0.52\linewidth]{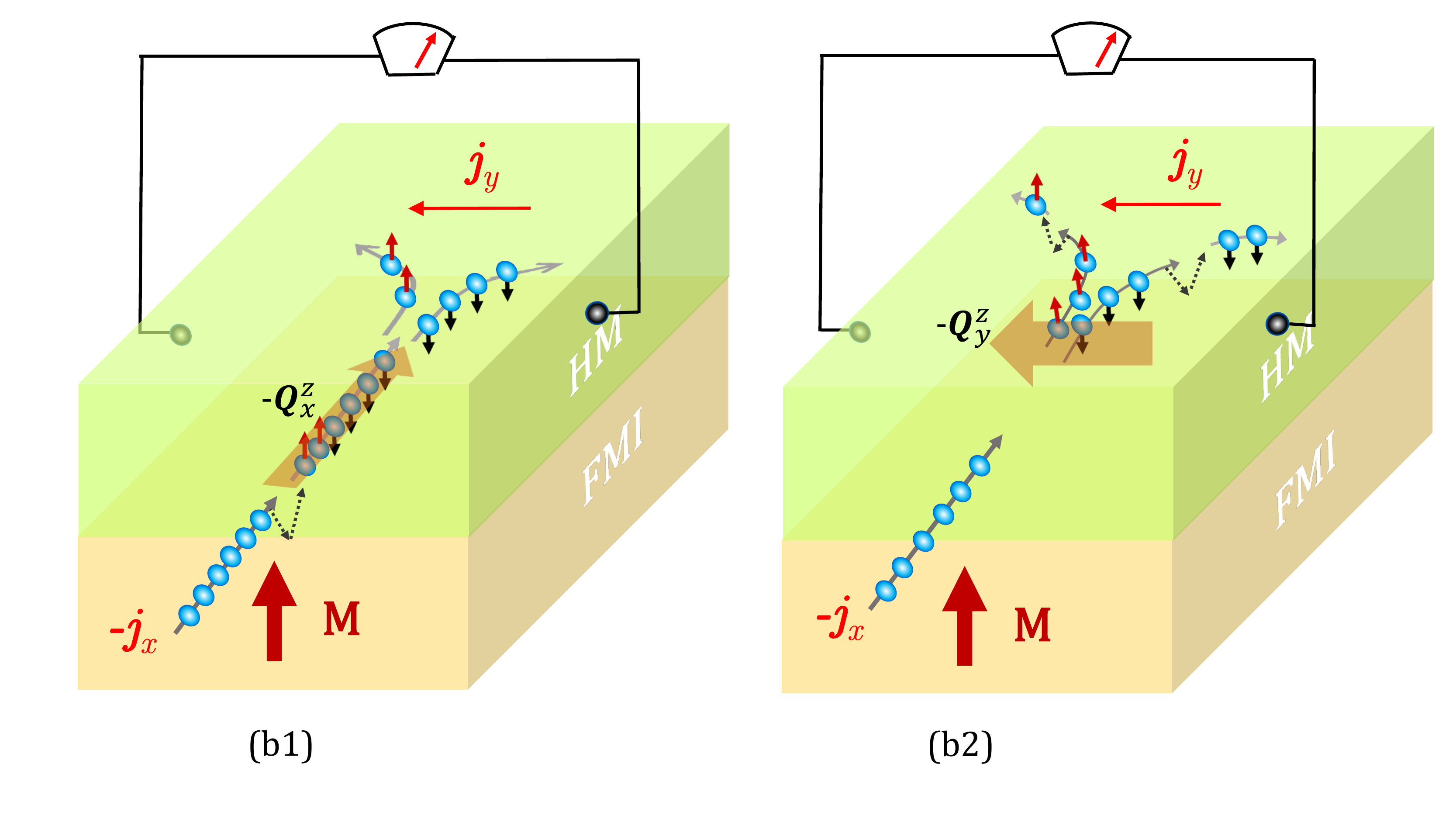}}
\hspace*{\fill}
\caption{Schematics of two different mechanisms of the AH effect in
heavy-metal (HM)/ferromagnetic-insulator (FMI) bilayers: (a) the spin Hall
AH mechanism and (b) nonlocal AH mechanism with two coexisting physical
processes depicted separately in panels b1 and b2. The curved arrows
represent the trajectories of electrons upon spin orbital scattering and the
dotted arrows stand for spin dependent scattering at the magnetic interface.}
\label{fig:schematics}
\end{figure*}

Recently, an AH signal has also been detected in a Platinum (Pt) layer in
direct contact with a YIG layer~\cite%
{Chien12PRL_Proximity-Pt,Althammer13PRB_SH-MR,Chien14PRL_AMR}. The former is
a non-magnetic heavy metal with strong spin orbit coupling and the latter is
a well-known ferromagnetic insulator. In view of the two aforementioned
ingredients for the AH effect in ferromagnets, it is puzzling that an AH
current would arise in Pt in the absence of spin polarized conduction
electrons. In a first attempt to solve the puzzle, Huang et. al.~\cite%
{Chien12PRL_Proximity-Pt} showed that the Pt layer in close proximity with
YIG acquires ferromagnetic characteristics, which essentially subsumes the
novel AH effect under the conventional AH effect for ferromagnetic metals.
This explanation ran into difficulties when it was found that the AH effect
persists in Pt/Cu/YIG trilayers~\cite{Saitoh13PRL_SH-MR} where the Cu layer
is deliberately inserted to eliminate the magnetic proximity effect.

An alternative explanation was then proposed~\cite%
{Saitoh13PRL_SH-MR,Bauer13PRB_SH-MR}, based on the physical mechanism
depicted in panel (a) of Fig.~\ref{fig:schematics}. In this mechanism the
applied charge current $j_{x}$ generates, via the spin Hall effect, a spin
current $Q_{z}^{y}$ propagating in the $z-$direction with spin along the $y-$
direction. When those electrons carrying $Q_{z}^{y}$ are reflected back from
the magnetic interface, spin rotation occurs and gives rise to a spin
current of $Q_{z}^{x}$, which in turn induces a transverse charge current $%
j_{y}$ via the inverse spin Hall effect~\cite%
{Saitoh06APL,Maekawa07PRL_inverse-SH}. Based on this picture, the transverse
electric current is of \textit{second order} in the spin orbit coupling or
spin Hall angle, which is qualitatively different from a conventional AH
current. It is worth mentioning that a fit to the experimental data based on
this model~\cite{Bauer13PRB_SH-MR,Althammer13PRB_SH-MR}, requires a spin
diffusion length on the order of $1$ $nm$. Such a short spin diffusion
length, an order of magnitude smaller than the room-temperature electron
mean path of Pt~\cite{Fischer80PRB_Pt-MFP}, casts doubt on the internal
consistency of the spin diffusion model.

In this paper, we propose a different mechanism for the AH current observed
in hybrid heavy-metal/ferromagnetic-insulator structures. The essential new
ingredient is the scattering of electrons from the (rough) metal-insulator
interface. Because the insulator is magnetic, the scattering rate is
spin-dependent (see Appendix B for a proof). This means that a charge
current flowing parallel to the interface is partially converted to a spin
current, while a spin current flowing parallel to the interface is partially
converted to a charge current. The surface-induced conversion of charge to
spin current and viceversa conspires with the spin Hall effect in the bulk
of the metal to produce the observed AH current. This may happen in two
ways: in the first process, (b1), the charge current $j_{x}$ generates, via
spin-dependent interfacial scattering a spin current $Q_{x}^{z}$, which
subsequently gives rise to the transverse spin polarized current $j_{y}$ via
the inverse spin Hall effect; in the second process, (b2), the applied
charge current $j_{x}$ first generates, via spin Hall effect, a transverse
spin current $Q_{y}^{z}$, which is then turned into a spin polarized current
$j_{y}$ due to spin dependent interfacial scattering. Both physical
processes involve spin orbit scattering only once (through the spin Hall
effect) and hence the resulting AH current is of \textit{first order} in the
spin orbit coupling or spin Hall angle. As a matter of fact, this AH effect
has the same physical nature as its conventional counterpart in bulk
ferromagnets, and differs from the latter only in the spatial separation of
the spin orbit interaction and the magnetization: it is for this reason that
we name it \textit{nonlocal AH effect}.

Compared to the double spin Hall effect mechanism proposed in Refs.~\cite%
{Saitoh13PRL_SH-MR,Bauer13PRB_SH-MR}, our proposal replaces one of the spin
Hall steps, the first or the second, by a spin-dependent interfacial
scattering. This leads to a good quantitative description of the transverse
current without the need of introducing an exceedingly small spin diffusion
length, as we will show in details in the remainder of the paper. In fact,
our mechanism survives in the limit of infinite spin diffusion length, while
the double spin Hall effect mechanism vanishes in that limit~\cite%
{Bauer13PRB_SH-MR}. In addition, the new mechanism has distinctive features
that can be tested experimentally, the most striking one being the sign of
the effect, which we predict to track the sign of the bulk spin Hall angle.

\textit{Linear response theory}$-$Let us consider a metal/insulator bilayer
as shown in Fig.~\ref{fig:schematics} with an external electric field
applied in the $x-$direction (i.e., $\mathbf{E}_{ext}=E_{ext}\hat{\mathbf{x}}
$) and with the magnetization of the insulator layer pointing in the $z$
direction, i.e., $\mathbf{m}=\hat{\mathbf{z}}$. We also assume that both
surfaces of the metal are rough, but on the average translational invariance
is recovered so that the transport properties are independent of $x$ and $y$
coordinates. The linear response of current densities to spin dependent
electric fields can be written as follows
\begin{eqnarray}
\mathbf{j}\left( z\right) &=&C_{0}\mathbf{E}\left( z\right) +C_{s}%
\boldsymbol{{\mathcal{E}}}_{\parallel }\left( z\right)  \notag \\
\mathbf{Q}_{\parallel }\left( z\right) &=&C_{0}\boldsymbol{{\mathcal{E}}}%
_{\parallel }\left( z\right) +C_{s}\mathbf{E}\left( z\right)  \notag \\
\mathbf{Q}_{\perp }\left( z\right) &=&C_{r}^{\prime }\boldsymbol{{\mathcal{E}%
}}_{\bot }\left( z\right) +C_{r}^{\prime \prime }\hat{\mathbf{z}}\times
\boldsymbol{{\mathcal{E}}}_{\bot }\left( z\right)
\label{Eq: LS-current-field}
\end{eqnarray}%
where $\mathbf{j}\left( z\right) =$ $(j_{x},j_{y})$ is the in-plane current
density (note that $j_{z}=0$ everywhere in the metal layer due to the open
boundary conditions), $\mathbf{Q}_{\parallel }=(Q_{x}^{z},Q_{y}^{z})$ is the
in-plane spin-current density (with spin in the $z$ direction), and $\mathbf{%
Q}_{\bot }=\left( Q_{z}^{x},Q_{z}^{y}\right) $ is the perpendicular-to-plane
spin current density with $Q_{z}^{x}$ and $Q_{z}^{y}$ carrying the $x$ and $%
y $ components of the spin. The corresponding fields are $\mathbf{E}%
=(E_{x},E_{y})$, $\boldsymbol{{\mathcal{E}}}_{\parallel }=(\mathcal{E}%
_{x}^{z},\mathcal{E}_{y}^{z})$ and $\boldsymbol{{\mathcal{E}}}_{\bot }=(%
\mathcal{E}_{z}^{x},\mathcal{E}_{z}^{y})$. Notice that $C_{k}$ is defined as
the integral operator with kernel $c_{k}\left( z,z^{\prime }\right) $, i.e.,
$C_{k}f\left( z\right) \equiv \int \mathrm{d}z^{\prime }c_{k}\left(
z,z^{\prime }\right) f\left( z^{\prime }\right) $. While $C_{0}$ is an
ordinary in-plane conductivity, $C_{s}$ describes the generation of an
in-plane spin current from an electric field in the presence of surface
scattering. As we show below, $C_{s}$ is the essential ingredient of our
theory, producing an AH current of first order in the spin Hall angle. On
the other hand, $C_{r}^{\prime }$ and $C_{r}^{\prime \prime }$ --
respectively the real and the imaginary part of the spin-mixing conductance~%
\cite{Brataas00PRL_finite-element} -- contribute only to second order. In
particular, $C_{r}^{\prime \prime }$ is the essential ingredient of the spin
Hall mechanism of the AH effect~\cite{Bauer13PRB_SH-MR}.

In the presence of the spin-orbit scattering, the driving electric fields $%
\mathbf{E},\mbox{\boldmath ${\cal E}$}_{\parallel },%
\mbox{\boldmath ${\cal
E}$}_{\perp }$ are self-consistently determined by the internal current
densities as follows
\begin{eqnarray}
\mathbf{E} &=&\mathbf{E}_{ext}+\rho _{sh}\hat{\mathbf{z}}\times (\mathbf{Q}%
_{\parallel }-\mathbf{Q}_{\perp })  \notag \\
&&\boldsymbol{{\mathcal{E}}}_{\parallel }=\rho _{sh}\hat{\mathbf{z}}\times
\mathbf{j}  \notag \\
&&\boldsymbol{{\mathcal{E}}}_{\bot }=-\rho _{sh}\hat{\mathbf{z}}\times
\mathbf{j}  \label{Eq: field-currents}
\end{eqnarray}%
where $\rho _{sh}\equiv \rho _{0}\theta _{sh}$ with $\rho _{0}$ being the
Drude resistivity and $\theta _{sh}$ the spin Hall angle of the metal layer.
Solving the system of linear equations~(\ref{Eq: LS-current-field}) and~(\ref%
{Eq: field-currents}), we obtain a general expression for the AH current
density up to $O\left( \theta _{sh}^{2}\right) $
\begin{equation}
j_{y}\left( z\right) =\left[ \rho _{sh}\left\{ C_{0},C_{s}\right\} -\rho
_{sh}^{2}C_{0}C_{r}^{\prime \prime }C_{0}\right] E_{ext}  \label{Eq:j_y(LR)}
\end{equation}%
where $\left\{ ,\right\} $ represents the anticommutator of the two integral
operators. 
Note that with finite $C_{s}$ the AH effect appears already at the first
order of $\theta _{sh}$. The two orderings of $C_{0}$ and $C_{s}$ in the
anticommutator of Eq.~(\ref{Eq:j_y(LR)}) correspond to the processes b1 and
b2 of Fig.~\ref{fig:schematics}. The second term on the right hand side of
Eq.~(\ref{Eq:j_y(LR)}) corresponds to the spin Hall AH effect which is of
second order in $\theta _{sh}$ and is proportional to the imaginary part of
the spin-mixing conductivity kernel. In what follows, we employ the
Boltzmann transport theory to explicitly construct the integral kernels $%
C_{0}$ and $C_{s}$ in the presence of a rough magnetic interface.

\textit{Boltzmann theory}$-$To quantitatively describe the nonlocal AH
effect in a heavy metal thin layer with an external electric field applied
in the $x-$direction (see Fig.~\ref{fig:schematics}(b)), we make use of the
spinor Boltzmann equation in the relaxation time approximation~\cite%
{sZhang00PRL,Valet93PRB,jwZhang04PRL,yQi2003PRB}
\begin{widetext}
\begin{equation}
v_{z}\frac{\partial \hat{f}\left( \boldsymbol{k},z\right) }{\partial z}%
-eE_{ext}v_{x}\left( \frac{\partial \hat{f}_{0}}{\partial \varepsilon _{k}}%
\right) +\frac{\boldsymbol{\sigma }\mathbf{\cdot }\left[ \boldsymbol{e}_{k}%
\mathbf{\times }\boldsymbol{\hat{\iota}}\left( k,z\right) \right]
}{\tau _{so}}=-\frac{\hat{f}\left( \boldsymbol{k},z\right)
-\hat{\bar{f}}\left( k,z\right) }{\tau }+\frac{2\hat{\bar{f}}\left(
k,z\right) -\hat{I}Tr_{\sigma }\hat{\bar{f}}\left( k,z\right) }{\tau
_{sf}}  \label{Eq: spinor-BTE}
\end{equation}%
\end{widetext}where $\hat{f}_{0}$ and $\hat{f}\left( \boldsymbol{k}\mathbf{,}%
z\right) $ are $2\times 2$ matrices represent respectively the equilibrium
and nonequilibrium spinor distribution functions, $\boldsymbol{v=}\mathrm{d}%
\varepsilon _{k}/\hbar \mathrm{d}\boldsymbol{k}$ is conduction electron
velocity, $\hat{\bar{f}}\left( k,z\right) \equiv \left( 1/4\pi \right) \int
\mathrm{d}\Omega _{\boldsymbol{k}}\hat{f}\left( \boldsymbol{k}\mathbf{,}%
z\right) $ is the angular average of the distribution and $\boldsymbol{\hat{%
\iota}}\left( k,z\right) \mathbf{\equiv }\left( 1/4\pi \right) \int \mathrm{d%
}\Omega _{\boldsymbol{k}}\mathbf{e}_{k}\hat{f}\left( \boldsymbol{k}\mathbf{,}%
z\right) $ is its dipolar moment, with $\boldsymbol{e}_{k}$ the unit vector
of $\boldsymbol{k}$. Non-spin-flip and spin-slip processes are included,
with $\tau $ and $\tau _{sf}$ being the momentum and spin relaxation times
respectively. The additional source term $\tau _{so}^{-1}\boldsymbol{\sigma }%
\mathbf{\cdot }\left[ \boldsymbol{e}_{k}\mathbf{\times }\boldsymbol{\hat{%
\iota}}\left( k,z\right) \right]$, where $\tau _{so}^{-1}$ is the spin-orbit
scattering rate, is responsible for the spin Hall effect~\cite%
{gVignale06PRB,Holstein72PRL,Halperin03PRB_2DEG-SOI}. It is this term that
generates the current-dependent fields in Eq.~(\ref{Eq: field-currents}).

The crucial step in our theory is the description of spin-dependent
interfacial scattering via boundary conditions for the distribution
function. For the interface (at $z=0$) between the heavy metal and the
ferromagnetic insulator, we impose the following generalized
Fuchs-Sondheimer boundary condition~\cite{slzhang15PRB_AMR},
\begin{equation}
\hat{f}^{+}(\boldsymbol{k}\mathbf{,}0)=\frac{1}{2}\hat{s}\hat{R}^{\dag }\hat{%
f}^{-}(\boldsymbol{k}\mathbf{,}0)\hat{R}+\frac{1}{2}\left( \hat{I}-\hat{s}%
\right) \left\langle \hat{f}^{-}(\boldsymbol{k}\mathbf{,}0)\right\rangle
+h.c.  \label{Eq:BC-z=0}
\end{equation}%
where $h.c.$ represents hermitian conjugate which ensures $\hat{f}^{+}$ to
be an hermitian, $\hat{I}$ is the $2\times 2$ identity matrix, $\left\langle
\hat{f}\right\rangle =\left( 2\pi \right) ^{-1}\int \mathrm{d}\phi _{%
\boldsymbol{k}}\hat{f}$ with $\phi _{\boldsymbol{k}}$ the $\boldsymbol{k}$%
-space azimuthal angle, and both $\hat{s}$ and $\hat{R}$ are $2\times 2$
matrices in spin space which are responsible for spin dependent specular
reflection and spin rotation of incident electrons.

The matrix $\hat{R}$, satisfying $\hat{R}^{\dagger }\hat{R}=\hat{I}$, is the
reflection amplitude matrix which captures the spin rotation of electrons
that are \textit{specularly reflected} from the magnetic interface (Note
that we assume such a coherent spin rotation does not occur for the
diffusively scattered electrons). The explicit form of $\hat{R}$ can be
determined by electron wave function matching subject to the following
\textit{spin-dependent} potential barrier
\begin{equation}
\hat{V}\left( z\right) =\left( V_{b}\hat{I}-J_{ex}\hat{\sigma}_{z}\right)
\Theta \left( -z\right)  \label{Eq:V^hat(z)}
\end{equation}%
where $V_{b}$ is the averaged potential barrier of the insulator, $J_{ex}$
measures the spin splitting of the energy barrier, $\hat{\sigma}_{z}$ is the
$z-$component of the Pauli spin matrices, $\ $and $\Theta \left( z\right) $
is the unit step function. Explicitly, $\hat{R}$ takes the following form
(see Appendix B for the derivation)
\begin{equation}
\hat{R}=\left( \frac{R^{\uparrow }+R^{\downarrow }}{2}\right) \hat{I}+\left(
\frac{R^{\uparrow }-R^{\downarrow }}{2}\right) \hat{\sigma}_{z}
\end{equation}%
where $R^{\sigma }=-\left( \kappa ^{\sigma }+ik_{z}\right) /\left( \kappa
^{\sigma }-ik_{z}\right) $ with $k_{z}$ the $z-$component of the electron
wave vector, $\kappa ^{\sigma }\equiv \sqrt{2m_{e}^{\ast }\left(
V_{b}-\sigma J_{ex}\right) -k_{z}^{2}}$ (we have let $\hbar =1$ for notation
convenience) and $m_{e}^{\ast }$ being the electron effective mass.

The matrix $\hat{s}$, on the other hand, is introduced to describe the
averaged effects of spin dependent scattering at the magnetic interface due
to roughness, impurities, etc. In general, we write~\cite%
{camley89PRL,Falicov94PRB}
\begin{equation}
\hat{s}=s_{0}\left( \hat{I}+p_{s}\hat{\sigma}_{z}\right)  \label{Eq:hat^s}
\end{equation}%
where $s_{0}\equiv \left( s^{\uparrow }+s^{\downarrow }\right) /2$ is the
average of the specular reflection coefficients $s^{\uparrow }$ and $%
s^{\downarrow }$ for spin-up and spin-down electrons with \textquotedblleft
up" and \textquotedblleft down" defined with respect to $\mathbf{m}$ ($=\hat{%
\mathbf{z}}$), and $p_{s}\equiv \left( s^{\uparrow }-s^{\downarrow }\right)
/\left( s^{\uparrow }+s^{\downarrow }\right) $ is their asymmetry. A simple
model calculation for the rough interface yields (see Appendix B for the
detailed calculation), to the lowest order in $J_{ex}/V_{b}$, the specular
reflection asymmetry $p_{s}\simeq -\frac{2J_{ex}}{V_{b}}\left(
1-s_{0}\right) $ for $s_{0}\lesssim 1$. Note that $p_{s}$ is \textit{negative%
}, meaning that more spin-down electrons are specularly scattered than
spin-up electrons, for the former encounter a higher energy barrier. Also,
we notice that a rough magnetic interface is essential for the spin
asymmetry of the specular reflection coefficients: for an ideally flat
interface, both $s^{\uparrow }$ and $s^{\downarrow }$ are exactly equal to
one, and no charge/spin conversion can occur.

For the outer surface at $z=d$, we assume, for simplicity, that the
scattering is diffusive, i.e.,%
\begin{equation}
\hat{f}^{-}(\boldsymbol{k}\mathbf{,}d)=\left\langle \hat{f}^{+}(\boldsymbol{k%
}\mathbf{,}d)\right\rangle  \label{Eq:BC-z=d}
\end{equation}%
Note that the boundary conditions given by Eqs.~(\ref{Eq:BC-z=0}) and (\ref%
{Eq:BC-z=d}) demand that both charge and spin currents flowing along the $z$%
-direction vanish at the outer (non magnetic) surface, whereas only the
charge current and the $z$-component of the spin current flowing along the $%
z $-direction vanish at the magnetic surface.

By solving the Boltzmann equation (\ref{Eq: spinor-BTE}) with the boundary
conditions given by Eqs.~(\ref{Eq:BC-z=0}) and (\ref{Eq:BC-z=d}), we have
calculated the current densities in the heavy-metal layer. Up to first order
in $\theta _{sh}$($\equiv \tau /\tau _{so}$), the Hall current density can
be expressed as follows
\begin{equation}
j_{y}^{ah}\left( z\right) =\rho _{sh}E_{ext}\int_{0}^{d}\frac{\mathrm{d}%
z^{\prime }}{l_{e}}\left[ c_{s}\left( z,z^{\prime }\right) \bar{c}_{0}\left(
z^{\prime }\right) +c_{0}\left( z,z^{\prime }\right) \bar{c}_{s}\left(
z^{\prime }\right) \right]  \label{Eq:j_y(z)_general}
\end{equation}%
where $l_{e}$ is the electron mean free path, the nonlocal integral kernels $%
c_{s}\left( z,z^{\prime }\right) $ and $c_{0}\left( z,z^{\prime }\right) $
are given by
\begin{equation}
c_{0}\left( z,z^{\prime }\right) =\frac{3}{4}\int_{0}^{1}\mathrm{d}\xi
\left( \xi ^{-1}-\xi \right) \left( s_{0}e^{-\frac{z+z^{\prime }}{l_{e}\xi }%
}+e^{-\frac{\left\vert z-z^{\prime }\right\vert }{l_{e}\xi }}\right)
\label{Eq: Theta(z,z')}
\end{equation}%
and
\begin{equation}
c_{s}\left( z,z^{\prime }\right) =\frac{3}{4}p_{s}\int_{0}^{1}\mathrm{d}\xi
\left( \xi ^{-1}-\xi \right) s_{0}e^{-\frac{z+z^{\prime }}{l_{e}\xi }}
\label{Eq: P(z,z')}
\end{equation}%
with their spatial averages defined as $\bar{c}_{0}\left( z\right) \equiv
\int_{0}^{d}\frac{\mathrm{d}z^{\prime }}{l_{e}}c_{0}\left( z,z^{\prime
}\right) $ and $\bar{c}_{s}\left( z\right) \equiv \int_{0}^{d}\frac{\mathrm{d%
}z^{\prime }}{l_{e}}c_{s}\left( z,z^{\prime }\right) $. The nonlocality of
the AH effect, i.e., the spatial separation of the spin-orbit scattering and
the magnetization, is clearly reflected in the structure of these integral
kernels which depend on the relative distance between the current and field
points as well as the distance of their center of mass coordinate from the
interface. Equations (\ref{Eq:j_y(z)_general})-(\ref{Eq: P(z,z')}) are the
main results of this paper.

One of the most remarkable features of the nonlocal AH effect is that it
appears at the first order of the spin Hall angle, which is distinctly
different from the spin Hall AH effect which occurs at the second order.
Since $p_{s}$ is negative, the directions of the nonlocal AH and the spin
Hall AH currents would be the \emph{same} for positive $\theta _{sh}$ but
the \emph{opposite} for negative $\theta _{sh}$, as can be seen from Eq.~(%
\ref{Eq:j_y(LR)}). Furthermore, the nonlocal AH is independent of spin
diffusion and thus is present in both ballistic and diffusive regimes,
whereas the spin Hall AH effect vanishes as the thickness of the metal layer
becomes much smaller than the spin diffusion length~\cite{Bauer13PRB_SH-MR}.

\begin{figure}[tbp]
\includegraphics[width=.45\textwidth]{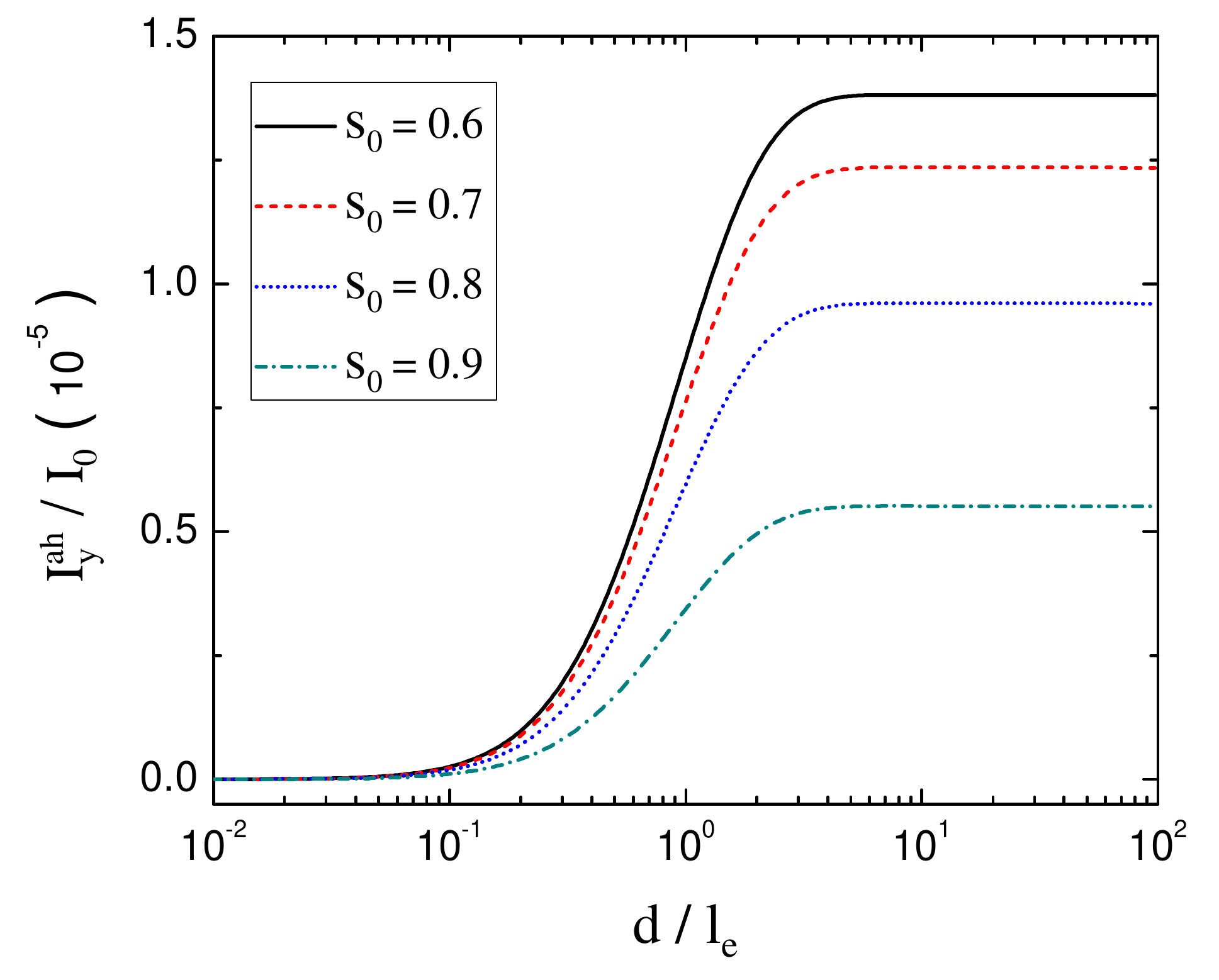}
\caption{The ratio of total AH current $I_{y}^{ah}$ to $I_{0}$ ($%
=c_{0}E_{ext}wd$) as a function of the thickness of the heavy metal layer
for several specular reflection parameters. Other Parameters: $\protect%
\theta _{sh}=0.05$, $J_{ex}=0.01$ $eV$ and $V_{b}=12$ $eV$.}
\label{fig:I_y-vs-d}
\end{figure}

The total AH current can be calculated from Eq.~(\ref{Eq:j_y(z)_general}) by
integrating the AH current density over the thickness of the layer, i.e., $%
I_{y}^{ah}\left( d\right) \equiv w\int_{0}^{d}\mathrm{d}zj_{y}^{ah}\left(
z\right) $ with $w$ being the width of the metal bar. By doing so, we find $%
I_{y}^{ah}\left( d\right) =2\rho _{sh}E_{ext}w\int_{0}^{d}\frac{\mathrm{d}%
z^{\prime }}{l_{e}}\bar{c}_{s}\left( z^{\prime }\right) \bar{c}_{0}\left(
z^{\prime }\right) $ where the factor of $2$ shows that the two physical
processes that we described in Fig.~\ref{fig:schematics}b contribute \emph{%
equally} to the total AH current. In Fig.~\ref{fig:I_y-vs-d}, we show the
thickness dependence of the total AH current for several values of the
specular reflection coefficient. We find that $I_{y}^{ah}$ begins to
saturate when the thickness reaches the electron mean free path. Also, we
note that the saturation current is smaller for a smoother surface (larger $%
s_{0}$), as expected from the above discussions.

\begin{figure}[tbp]
\includegraphics[width=.45\textwidth]{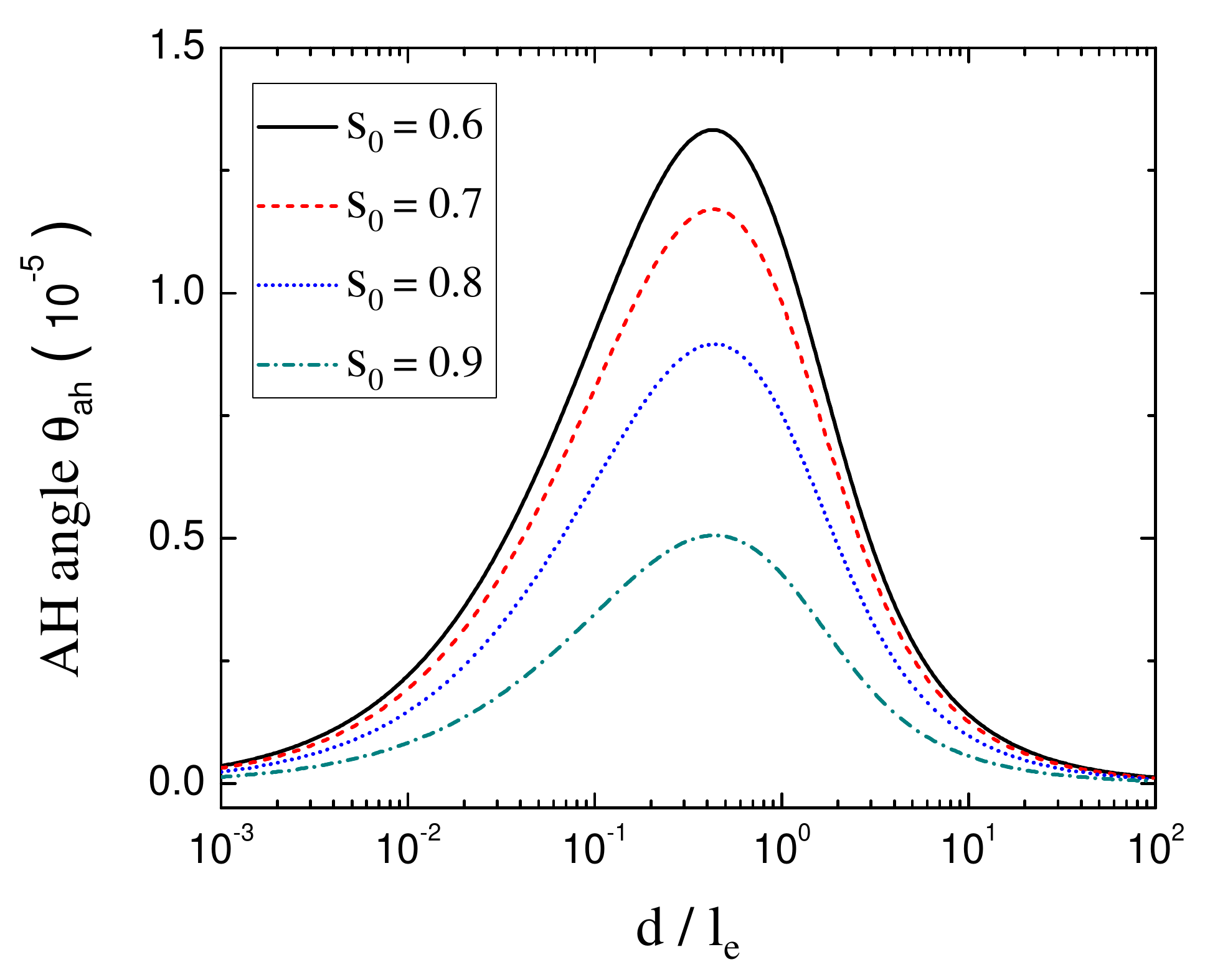}
\caption{The AH angle $\protect\theta_{ah}$ ($\equiv \bar{\protect\rho}%
_{xy}^{ah}\left( d\right) /\bar{\protect\rho}_{xx}\left( d\right)$) as a
function of thickness of the heavy metal layer for several values of the
specular reflection coefficient. Other Parameters: $\protect\theta %
_{sh}=0.05 $, $J_{ex}=0.01$ $eV$ and $V_{b}=12$ $eV$.}
\label{fig:theta_ah-vs-d}
\end{figure}

Experimentally, a most relevant quantity is the ratio of the spatially
averaged AH resistivity to the longitudinal resistivity, i.e., $\theta
_{ah}\equiv \bar{\rho}_{xy}^{ah}\left( d\right) /\bar{\rho}_{xx}\left(
d\right) $. The AH resistivity can be obtained by inverting the conductivity
tensor. Since $p_{s}s_{0}\theta _{sh}$$\lesssim 10^{-1}$, to a good
approximation, we can take $\bar{\rho}_{xy}^{ah}\simeq \bar{c}_{xy}^{ah}/%
\bar{c}_{xx}^{2}$ where $\bar{c}_{xy}^{ah}\equiv d^{-1}\int_{0}^{d}\mathrm{d}%
zj_{y}^{ah}\left( z\right) /E_{ext}$ with $j_{y}^{ah}\left( z\right) $ given
by Eq.~(\ref{Eq:j_y(z)_general}). In Fig.~\ref{fig:theta_ah-vs-d}, we show
the thickness dependence of $\theta _{ah}$ for several values of the
specular reflection coefficient $s_{0}$. For $d\ll l_{e}$, $\theta _{ah}$
tends to zero, because $\bar{\rho}_{xx}\left( d\right) $ increases with
decreasing layer thickness. In the opposite limit of $d\gg l_{e}$, $\theta
_{ah}$ also diminishes since the nonlocal AH effect is essentially an
interface effect, which saturates for thicknesses larger than the electron
mean path. By choosing the following parameters for a Pt ($7$ $nm$)/YIG
bilayer at room temperature: $\theta _{sh}=0.05$~\cite{lqLiu11arXiv_SH}, $%
s_{0}=0.6$, $J_{ex}=0.01$ $eV$~\cite{Kajiwara10Nature}, $V_{b}=12$ $eV$and $%
l_{e}=20$ $nm$\cite{Fischer80PRB_Pt-MFP}, we estimate the AH angle arising
from our mechanism to be about $1.3\times 10^{-5}$, which is in good
agreement with experimental observations~\cite%
{Chien12PRL_Proximity-Pt,Althammer13PRB_SH-MR}.

As a final point, we suggest a crucial verification of our mechanism by
contrasting the directions of the Hall current (or the signs of Hall
voltages) of two trilayer structures Pt/Cu/YIG and $\beta $-Ta/Cu/YIG. Since
the spin Hall angles of Pt and $\beta $-Ta~are of opposite signs~\cite%
{lqLiu12Science,Morota11PRB_SH-Ta-Pt,Hahn13PRB_SMR-Pt-Ta} we predict that
the Hall current directions in these two trilayers will be opposite. A Cu
layer, thinner than the electron mean free path, may be inserted between the
heavy-metal and the magnetic insulator in order to eliminate the magnetic
proximity effect, while the nonlocal AH effect will still be operative.

\emph{Acknowledgement.}$-$It is a pleasure to thank O. Heinonen, S. Zhang,
A. Hoffmann, W. Jiang and W. Zhang for various stimulating discussions. One
of the author, S. S.-L. Zhang is deeply indebted to O. Heinonen for his
hospitality at Argonne National Lab, where part of the work was done. This
work was supported by NSF Grants DMR-1406568.

\appendix

\section{Appendix A: Spinor reflection amplitude at a
metal/magnetic-insulator interface}

Consider the following free electron Hamiltonian for a
metal/magnetic-insulator interface
\begin{equation}
\hat{H}=\frac{\hat{\mathbf{p}}^{2}}{2m_{e}^{\ast }}+\left( V_{b}\hat{I}%
-J_{ex}\hat{\sigma}_{z}\right) \Theta \left( -z\right)  \tag{A1}
\label{Eq: Hamiltonian}
\end{equation}%
where $V_{b}$ is the spin-averaged barrier for electrons to go from the
metal to the insulator, $J_{ex}$ is the exchange coupling which is
responsible for the spin-splitting of energy levels in the insulator, and $%
\Theta \left( z\right) $ is the unit step function. Here we have chosen the
spin quantization axis to be parallel to the magnetization $\mathbf{m}$ ($=%
\hat{\mathbf{z}}$). For an incident electron (from the metal side, $z>0$)
with its spin pointing in the direction $\left( \theta ,\phi \right) $ with
respect to $\mathbf{m}$, we can write the scattering wave function as
follows
\begin{equation}
\hat{\psi}\left( \mathbf{r}\right) =\cos \left( \frac{\theta }{2}\right)
e^{-i\phi /2}\varphi ^{\uparrow }\left( \mathbf{r}\right) \left\vert
\uparrow \right\rangle +\sin \left( \frac{\theta }{2}\right) e^{i\phi
/2}\varphi ^{\downarrow }\left( \mathbf{r}\right) \left\vert \downarrow
\right\rangle  \tag{A2}
\end{equation}%
where the spatial parts of the spinor wave function are
\begin{equation}
\varphi ^{\sigma }\left( \mathbf{r}\right) =\left\{
\begin{array}{c}
\left( e^{-ik_{z}z}+R^{\sigma }e^{ik_{z}z}\right) e^{i\mathbf{q\cdot }%
\boldsymbol{\rho }}\text{, \ \ }z>0 \\
T^{\sigma }e^{\kappa ^{\sigma }z}e^{i\mathbf{q\cdot }\boldsymbol{\rho }}%
\text{, \ \ \ \ \ \ \ \ \ \ \ \ \ \ \ \ \ }z<0%
\end{array}%
\right.  \tag{A3}
\end{equation}%
where $\sigma =\uparrow \left( \downarrow \right) $, $R^{\sigma }$ and $%
T^{\sigma }$ are the corresponding reflection and transmission amplitudes, $%
\mathbf{k=}\left( \mathbf{q,}k_{z}\right) $ and $\mathbf{r=}\left(
\boldsymbol{\rho },z\right) $ are the wave vector and spatial coordinates
respectively, and $\kappa ^{\sigma }=\sqrt{k_{b}^{2}-\sigma
k_{J}^{2}-k_{z}^{2}}$ with $k_{b}\equiv \sqrt{2m_{e}^{\ast }V_{b}/\hbar ^{2}}
$ and $k_{J}\equiv \sqrt{2m_{e}^{\ast }J_{ex}/\hbar ^{2}}$. By matching the
wave functions and their derivatives at $z=0$, we find
\begin{equation}
R^{\sigma }=-\frac{\kappa ^{\sigma }+ik_{z}}{\kappa ^{\sigma }-ik_{z}}
\tag{A4}  \label{Eq: R^sigma}
\end{equation}%
and
\begin{equation}
T^{\sigma }=1+R^{\sigma }=-\frac{2ik_{z}}{\kappa ^{\sigma }-ik_{z}}\text{.}
\tag{A5}
\end{equation}

\section{Appendix B: Spin dependent specular reflection coefficient}

In this section, we prove that the specular reflection coefficient $s$ is
spin-dependent for a rough metal/magnetic-insulator interface. In the
absence of interface roughness, the bilayer can be modeled as a simple
spin-dependent step potential as given in Eq.~(\ref{Eq: Hamiltonian}), the
corresponding free electron Green's function (setting $\hbar =1$) reads
\begin{equation}
g_{\mathbf{q}}^{\sigma }(z,z^{\prime };E)=\frac{m_{e}^{\ast }}{ik_{z}}\left[
e^{ik_{z}|z-z^{\prime }|}+R^{\sigma }e^{-ik_{z}(z+z^{\prime })}\right] \,
\tag{B1}  \label{Eq: g_q^sigma}
\end{equation}%
where $z<0$ and $z^{\prime }<0$, $k_{z}=\sqrt{2m_{e}^{\ast }E-q^{2}}$ with $%
E $ the total kinetic energy and $\mathbf{q}$ the in-plane momentum, and the
reflection amplitude for electron with spin $\sigma $ is given by Eq.~(\ref%
{Eq: R^sigma}).

Now we model a rough interface by a set of randomly-distributed impurities
localized at the interface ($z=0$) with $\delta $-correlated potential $%
V_{imp}(\mathbf{r})$ satisfying the following properties~\cite%
{xgZhang95PRB,Stewart03PRB,Los05PRB_surfRough}
\begin{equation}
\left\langle V_{imp}\left( \mathbf{r}\right) \right\rangle =0  \tag{B2}
\end{equation}%
and
\begin{equation}
\left\langle V_{imp}\left( \mathbf{r}\right) V_{imp}\left( \mathbf{r}%
^{\prime }\right) \right\rangle =\gamma \delta \left( \boldsymbol{\rho }-%
\boldsymbol{\rho }^{\prime }\right) \delta \left( z\right) \delta \left(
z^{\prime }\right)  \tag{B3}
\end{equation}%
where $\left\langle {}\right\rangle $ denotes the impurity ensemble average
and $\gamma $ describes the amplitude of the fluctuation. Up to first order
in $\gamma $, the \textit{impurity-averaged} Green's function reads~\cite%
{Stewart03PRB}
\begin{align}
\left\langle G_{\mathbf{q}}^{\sigma }(z,z^{\prime };E)\right\rangle & =g_{%
\mathbf{q}}^{\sigma }(z,z^{\prime };E)  \notag \\
& +\gamma g_{\mathbf{q}}^{\sigma }(z,0;E)g_{\mathbf{q}}^{\sigma
}(0,z^{\prime };E)N^{\sigma }(0;E)  \tag{B4}  \label{Eq: ave-G_q^Sigma}
\end{align}%
where $N^{\sigma }(0;E)\equiv $ $\int \frac{\mathrm{d}\mathbf{q}^{\prime }}{%
\left( 2\pi \right) ^{2}}g_{\mathbf{q}^{\prime }}^{\sigma }(0,0;E)$. By
placing Eq.~(\ref{Eq: g_q^sigma}) into Eq.~(\ref{Eq: ave-G_q^Sigma}), we
find
\begin{align}
\left\langle G_{\mathbf{q}}^{\sigma }(z,z^{\prime };E)\right\rangle & =\frac{%
m}{ik_{z}}\left\{ e^{ik_{z}|z-z^{\prime }|}\right.  \notag \\
& \left. +e^{-ik_{z}(z+z^{\prime })}R^{\sigma }\left[ 1-2i\gamma N^{\sigma
}(0;E)\frac{k_{z}}{U_{b}^{\sigma }}\right] \right\}  \tag{B5}
\label{Eq: ave-G_v3}
\end{align}%
where $U_{b}^{\sigma }\equiv V_{b}-\sigma J_{ex}$ is the spin-dependent
barrier. Comparing Eq.~(\ref{Eq: ave-G_v3}) with Eq.~(\ref{Eq: g_q^sigma}),
we identify the effective reflection amplitude in the presence of the
surface roughness as
\begin{equation}
\bar{R}^{\sigma }=R^{\sigma }\left[ 1-2i\gamma N^{\sigma }(0;E)\frac{k_{z}}{%
U_{b}^{\sigma }}\right]  \tag{B6}  \label{Eq: R_bar}
\end{equation}%
Up to $O\left( \gamma \right) $, the reflection coefficient is
\begin{equation}
r^{\sigma }=\left\vert R^{\sigma }\right\vert ^{2}\left[ 1-2\gamma A^{\sigma
}\left( 0;E\right) \frac{k_{z}}{U_{b}^{\sigma }}\right]  \tag{B7}
\end{equation}%
with the surface spectral function defined as\textbf{\ }$A^{\sigma }\left(
0;E\right) =-2\Im m~N^{\sigma }(0;E)$. We thus identify the specular
reflection coefficient as
\begin{equation}
s^{\sigma }=1-2\gamma A^{\sigma }\left( 0;E\right) \frac{k_{z}}{%
U_{b}^{\sigma }}  \tag{B8}  \label{Eq: s^sigma_withA}
\end{equation}%
By placing Eq.~(\ref{Eq: g_q^sigma}) into Eq.~(\ref{Eq: s^sigma_withA}) and
carrying out the integration over in-plane momentum $\mathbf{q}$, we obtain
an explicit expression for $s^{\sigma }$

\begin{equation}
s^{\sigma }=1-\gamma \frac{2k_{z}\left( 2m_{e}^{\ast }E\right) ^{3/2}}{3\pi
\left( U_{b}^{\sigma }\right) ^{2}}\simeq 1-\gamma \frac{2k_{z}k_{F}^{3}}{%
3\pi V_{b}^{2}}\left( 1+\sigma \frac{2J_{ex}}{V_{b}}\right)  \tag{B9}
\label{Eq: s^sigma_q-dep}
\end{equation}%
where we have replaced the total kinetic energy $E$ by the Fermi energy and
kept term up to $O\left( J_{ex}/V_{b}\right) $. Therefore, we have shown
that the specular reflection coefficient is indeed spin-dependent. We also
note that $s_{\sigma }$ is in general dependent on the \textit{direction} of
the incident momentum. For brevity, we shall work with an angle-averaged
specular reflection coefficient, i.e.,
\begin{equation}
\bar{s}_{\sigma }=\int \mathrm{d}\Omega _{\mathbf{k}}s_{\sigma }\left(
q\right) /4\pi =1-\frac{\gamma k_{F}^{4}}{3\pi ^{2}V_{b}^{2}}\left( 1+\sigma
\frac{2J_{ex}}{V_{b}}\right)  \tag{B10}
\end{equation}%
It follows that the spin averaged specular reflection coefficient as well as
the spin symmetry of the specular reflection can be expressed as (up to $%
O\left( J_{ex}/V_{b},\gamma \right) $)
\begin{equation}
s_{0}\equiv \frac{\bar{s}^{\uparrow }+\bar{s}^{\downarrow }}{2}=1-\gamma
\cdot \frac{k_{F}^{4}}{3\pi ^{2}V_{b}^{2}}  \tag{B11}  \label{Eq: s_0_final}
\end{equation}%
and
\begin{equation}
p_{s}\equiv \frac{\bar{s}^{\uparrow }-\bar{s}^{\downarrow }}{\bar{s}%
^{\uparrow }+\bar{s}^{\downarrow }}\simeq -\gamma \cdot \frac{J_{ex}}{V_{b}}%
\cdot \frac{2k_{F}^{4}}{3\pi ^{2}V_{b}^{2}}  \tag{B12}  \label{Eq: p_s_final}
\end{equation}%
Interestingly, we note the $p_{s}$ has a negative sign; in other words, the
specular reflection coefficient for spin-up electrons is smaller than that
of the spin-down electrons as the latter encounter a higher barrier.

Eliminating the parameter $\gamma $ from Eqs.~(\ref{Eq: s_0_final}) and (\ref%
{Eq: p_s_final}), we find an approximate relation between $s_{0}$ and $p_{s}$%
\begin{equation}
p_{s}\simeq -\frac{2J_{ex}}{V_{b}}\left( 1-s_{0}\right)   \tag{B13}
\end{equation}%
This relation is valid for a moderately rough interface, i.e., $%
s_{0}\lesssim 1$.

\bibliographystyle{my-asp-style}
\bibliography{20151011_nonlocal-anomalous-Hall}

\end{document}